%% file: main.tex
\documentclass[conference,table,xcdraw]{IEEEtran}
\usepackage{blindtext, graphicx}
\usepackage{amsthm}
\usepackage{tikz}
\usetikzlibrary{decorations.text,calc,fit}
\usepackage{adjustbox}
\usepackage{algpseudocode}
\usepackage{cite}
\usepackage{comment}
\usepackage{courier}
\usepackage{multicol}
\usepackage{multirow}
\usepackage{stfloats}
\usepackage{cite}

\usepackage{mathtools,amssymb,lipsum}

\usepackage{cuted}
\usepackage{amsfonts}
\usepackage{graphicx}
\usepackage{gensymb}
\usepackage{textcomp}
\usepackage{xcolor}
\def\BibTeX{{\rm B\kern-.05em{\sc i\kern-.025em b}\kern-.08em
    T\kern-.1667em\lower.7ex\hbox{E}\kern-.125emX}}

\usepackage{bm}
\usepackage{bbm}
\usepackage{subcaption}
\usepackage{float}
\usepackage{xfrac}
\usepackage{pdfpages}

\usepackage{stfloats}

\newcommand{\esp}[2]{\mathbb{E}_{#1}\left[#2\right]}

\renewcommand{\Pr}[2]{\mathbb{P}_{#1}\left(#2\right)}
\renewcommand{\exp}[1]{\mathrm{exp}\left(#1\right)}
\renewcommand{\Im}[1]{\mathrm{Im}\left[#1\right]}

\DeclareMathOperator*{\argmin}{argmin}

\begin{document}

 \title{Stochastic Geometry-based Modelling of Mobile UAV Relay Networks under Realistic Fading}

\author{\IEEEauthorblockN{François De Saint Moulin\IEEEauthorrefmark{1}, Charles Wiame\IEEEauthorrefmark{2}, Claude Oestges\IEEEauthorrefmark{3}, Luc Vandendorpe\IEEEauthorrefmark{4}}
\IEEEauthorblockA{ICTEAM, UCLouvain - Louvain-la-Neuve, Belgium\\}
\IEEEauthorrefmark{1}francois.desaintmoulin@uclouvain.be, \IEEEauthorrefmark{2}charles.wiame@uclouvain.be,\\ \IEEEauthorrefmark{3}claude.oestges@uclouvain.be,
\IEEEauthorrefmark{4}luc.vandendorpe@uclouvain.be}

\maketitle
\thispagestyle{plain}
\pagestyle{plain}

\begin{abstract}
We consider a relay network based on Unmanned Aerial Vehicles (UAV). Terrestrial Base Stations (TBS) and UAV Relay Nodes (RN) are modelled using two Homogeneous Poisson Point Processes (HPPP). UAVs can hover at a fixed position or move following specific displacement schemes. The Coverage Probability (CP) of a typical user equipment (UE) is derived, either when it communicates via a direct link (from the TBS to the UE) or via a relay link (from the TBS to the UE through a UAV RN). Every link can be in Line-of-Sight (LoS) or Non Line-of-Sight (NLoS), and suffers from Rician fading with distance-dependent parameters. This coverage is calculated by means of both stochastic geometry (SG) and Monte-Carlo (MC) simulations. The benefits of the use of UAV as RNs are analysed depending on their altitude, density, and mobility scheme.
\end{abstract}

\vspace*{0.1cm}



\section{Introduction}
\input{Introduction.tex}

\section{System Model}
\input{System_Model.tex}

\section{Coverage Probability}
\label{section:coverage_probability}
\input{Coverage_Probability.tex}

\section{Numerical Results}
\input{Numerical_Results.tex}

\section{Conclusion}
\label{section:conclusion}
\input{Conclusion.tex}

\bibliographystyle{ieeetr}
\bibliography{biblio}


\clearpage
\onecolumn
\appendix
\textbf{A. Proof of Lemma 1}\medskip

Conditioned on $r_{SD}$ and $r_{RD,0}$, the angle $\Theta$ between $\tilde{\bm{\phi}}_{SD}$ and $\tilde{\bm{\phi}}_{RD}(t)$ is uniformly distributed between 0 and $\pi$. Additionally, the distribution of $r_{SR}(t)$ is time-invariant and the Complementary Cumulative Distribution Function (CCDF) of $r_{SR,0}$ is given in Appendix B. Therefore, 
\begin{align*}
    \Pr{\bm{\tilde{\phi}}_{SD},\bm{\tilde{\phi}}_{SR}(t)|r_{SD},r_{RD,0}}{\bm{\tilde{\phi}}_{SD} \neq \bm{\tilde{\phi}}_{SR}(t)}    &=\Pr{r_{SR}(t),\Theta|r_{SD},r_{RD,0}}{r_{SR}(t)\leq \sqrt{r_{SD}^2+r_{RD}^2(t)-2r_{SD}r_{RD}(t) \cos\Theta}}\\
    &= \frac{1}{\pi}\int_0^\pi \left(1 -  \exp{-\pi\lambda_T\left(r_{SD}^2+r_{RD}^2(t)-2r_{SD}r_{RD}(t) \cos\theta\right)}\right) \mathrm{d}\theta\\
    &= 1-\exp{-\pi\lambda_T\left(r_{SD}^2+r_{RD}^2(t)\right)}I_0\left(2\pi\lambda_Tr_{SD}r_{RD}(t)\right).
\end{align*}
\smallskip

\textbf{B. Proof of Lemma 2}\medskip

The CCDFs of $r_{k,0}$ are obtained thanks to the void probability of a HPPP:
\begin{equation*}
    \Pr{r_{k,0}}{r_{k,0} > r} = \Pr{\Psi_{\bm{\Phi}_k}}{\Psi_{\bm{\Phi}_k}(\mathcal{B}(\tilde{\bm{\phi}}_{k,0},r)) = 0} \overset{(a)}{=} \Pr{\Psi_{\bm{\Phi}_k}}{\Psi_{\bm{\Phi}_k}(\mathcal{B}(0,r)) = 0} = \exp{-\pi\lambda_k r^2},
\end{equation*}
where $(a)$ is obtained thanks to the stationarity property of a HPPP. The function $\Psi_{\bm{\Phi}_k}$ is the random counting measure of $\bm{\Phi}_k$ and $\mathcal{B}(\bm{c},r)$ is a ball of radius $r$ centered on $\bm{c}$.

The PDFs of $r_{k,0}$ are then given by
\begin{equation*}
    f_{r_{k,0}}(r) = -\frac{\mathrm{d}}{\mathrm{d}r}\Pr{r_{k,0}}{r_{k,0} > r} = 2\pi\lambda_k\:r\:\exp{-\pi\lambda_k r^2}.
\end{equation*}
\smallskip

\textbf{C. Proof of Lemma 3}\medskip

Since $r_{SD}$ and $r_{RD}(t)$ are two independent random variables, using the PDFs of $r_{k,0}$ computed in Appendix B, the probability $\mathcal{A}_{SD}$ that the typical UE selects the direct link is given by
\begin{align*}
  \mathcal{A}_{SD} &= \Pr{r_{SD},r_{RD,0}}{r_{SD} \leq \sqrt{r_{RD}^2(t) + H_R^2}}\\
 &= \int_0^{vt} f_{r_{RD,0}}(r) \int_0^{H_R} f_{r_{SD}}(r')\mathrm{d}r'\mathrm{d}r + \int_{vt}^\infty f_{r_{RD,0}}(r) \int_0^{\sqrt{(r-vt)^2 + H_R^2}} f_{r_{SD}}(r')\mathrm{d}r'\mathrm{d}r\\
&= 1 - \exp{-\pi\lambda_T H_R^2}\left(1-\exp{-\pi\lambda_Rv^2t^2}\right)\\
&\ \ \ \ \ \ \ - 2\pi\lambda_R\:\exp{-\pi\lambda_T(H_R^2+v^2t^2)}\int_{vt}^\infty r\:\exp{-\pi(\lambda_R+\lambda_T)r^2 + 2\pi\lambda_T vt\:r}\mathrm{d}r.
\end{align*}
The probability $\mathcal{A}_{SRD}$ that the typical UE selects the relay link is then computed as $\mathcal{A}_{SRD} = 1 - \mathcal{A}_{SD}.$
\bigskip

\textbf{D. Proof of Proposition 1}\medskip

The CP of the typical UE at time $t \geq 0$ for a given SINR threshold $\beta$ is given by
\begin{align*}
\mathcal{P}(\beta,t) &= \Pr{\chi(t)}{\chi(t) \geq \beta} = \esp{r_{SD},r_{RD,0}}{\Pr{\chi(t)|r_{SD},r_{RD,0}}{\chi(t) \geq \beta}},
\end{align*}
where $\mathbb{E}_A$ denotes the expectation over the random variable $A$. Additionally, the SINR at the typical UE $\chi(t)$ is given by
\begin{equation*}
\chi(t) = \left\{
\begin{alignedat}{3}
&\chi_{SRD}(t) &&\:\mathrm{if}\: && r_{SD} >    \sqrt{r_{RD}^2(t) + H_R^2}\\
&\chi_{SD}  &&\:\mathrm{if}\: && r_{SD} \leq \sqrt{r_{RD}^2(t) + H_R^2}
\end{alignedat}
 \right. .
\end{equation*}
When the relay link is selected, with the DF protocol and Approximation 1, we have
\begin{align*}
\nonumber \mathcal{P}_{SRD|r_{SD},r_{RD,0}}(\beta,t) = \mathcal{P}_{SRD|r_{RD,0}}(\beta,t) &= \Pr{\chi_{SRD}(t)|r_{SD},r_{RD,0}}{\chi_{SRD}(t) \geq \beta} \\
&= \underbrace{\Pr{\chi_{SR}}{\chi_{SR} \geq \beta}}_{\mathcal{P}_{SR}(\beta)}\underbrace{\Pr{\chi_{RD}(t)|r_{RD,0}}{\chi_{RD}(t) \geq \beta}}_{\mathcal{P}_{RD|r_{RD,0}}(\beta,t)}.
\end{align*}
Therefore, the integration of the CP conditioned on $r_{SD}$ and $r_{RD,0}$ over the four regions of Figure \ref{fig:associationRegionsNNAMob} is computed as following:
\begin{align*}
\nonumber \mathcal{P}(\beta,t) &=\int_0^{H_R}\mathcal{P}_{SD|r_{SD}=r'}(\beta)f_{r_{SD}}(r')\int_0^\infty   f_{r_{RD}}(r)\mathrm{d}r\mathrm{d}r' + \int_{H_R}^\infty \mathcal{P}_{SD|r_{SD}=r'}(\beta)f_{r_{SD}}(r')\int_{\sqrt{r'^2-H_R^2}+vt}^\infty   f_{r_{RD}}(r)\mathrm{d}r\mathrm{d}r'\\
\nonumber &+ \int_{0}^{vt}\mathcal{P}_{SRD|r_{RD,0}=r}(\beta,t)f_{r_{RD}}(r)\int_{H_R}^\infty   f_{r_{SD}}(r')\mathrm{d}r'\mathrm{d}r + \int_{vt}^{\infty} \mathcal{P}_{SRD|r_{RD,0}=r}(\beta,t)f_{r_{RD}}(r)\int_{\sqrt{(r-vt)^2+H_R^2}}^\infty   f_{r_{SD}}(r')\mathrm{d}r'\mathrm{d}r\\
\nonumber &= \int_0^{H_R}\mathcal{P}_{SD|r_{SD}=r'}(\beta)f_{r_{SD}}(r')\mathrm{d}r' \\
&+ \int_{H_R}^\infty \mathcal{P}_{SD|r_{SD}=r'}(\beta) f_{r_{SD}}(r')\:\exp{-\pi\lambda_R \left(v^2t^2-H_R^2\right)}  \exp{-\pi\lambda_R \left(r'^2 + 2vt\sqrt{r'^2-H_R^2}\right)} \mathrm{d}r'\\
&+ \mathcal{P}_{SR}(\beta)\int_{0}^{vt} \mathcal{P}_{RD|r_{RD,0}=r}(\beta,t)f_{r_{RD}}(r)\:\exp{-\pi\lambda_T H_R^2}\mathrm{d}r \\
&+ \mathcal{P}_{SR}(\beta)\int_{vt}^\infty \mathcal{P}_{RD|r_{RD,0}=r}(\beta,t)f_{r_{RD}}(r)\:\exp{-\pi\lambda_T \left(H_R^2+v^2t^2\right)}\exp{-\pi\lambda_T (r^2-2vt\:r)}\mathrm{d}r.
\end{align*}
Finally, Equations \eqref{eqn:I1}-\eqref{eqn:I4} are obtained knowing that
\begin{equation*}
    \mathcal{P}_{SR}(\beta) = \int_0^\infty \mathcal{P}_{SR|r_{SR}(t)=r}(\beta) f_{r_{SR}(t)}(r)\:\mathrm{d}r.
\end{equation*}
\smallskip

\textbf{E. Proof of Proposition 2}\medskip

The CP of each link in LoS/NLoS conditioned on the distances between the two communicating nodes can be developed as
\begin{align*}
    \nonumber \mathcal{P}_{k|n,r_k(t)}(\beta,t) &= \Pr{\left|h_{k,n|r_k(t) }\right|^2,I_{k}(t)|r_{k}(t)}{\frac{P_k G_{k,M} L_{k,n}^{-1}(r_k(t)) \left|h_{k,n|r_k(t)}\right|^2}{I_{k}(t) + \sigma^2} \geq \beta}\\
   \nonumber &= \Pr{\left|h_{k,n|r_k(t)}\right|^2,I_k(t)|r_{k}(t)}{\underbrace{P_k G_{k,M}
     L_{k,n}^{-1}(r_k(t)) \left|h_{k,n|r_k(t)}\right|^2 - \beta I_{k}(t)}_{Z_{k|r_k(t)}(t)} \geq \beta\sigma^2}\\
    &= \frac{1}{2} + \frac{1}{\pi}\int_0^\infty\frac{1}{\tau}\Im{\phi_{Z_{k|r_k(t)}(t)}(\tau)\:\exp{-j\tau\beta \sigma^2}}\mathrm{d}\tau.
\end{align*}
Since $Z_{k|r_k(t)}(t)$ is a linear combination of $|h_{k,n|r_k(t)}|^2$ and $I_{k}(t)$ which are two independent random variables, knowing that $\phi_{aX}(\tau) = \phi_{X}(a\tau)$ for any random variable $X$ and $a\in \mathbb{R}$, the characteristic function of $Z_{k|r_{k}(t)}(t)$ can be expressed as
\begin{equation*}
    \phi_{Z_{k}(t)|r_{k}(t)}(\tau) = \phi_{|h_{k,n|r_k(t)}|^2}(\tau P_k G_{k,M} 
     L_{k,n}^{-1}(r_k(t))) \phi_{I_{k}(t)}(-\tau\beta).
\end{equation*}
Equation \eqref{eqn:PSDRSD} is finally obtained using the relation $\phi_X(\tau) = \mathcal{L}_X(-j\tau)$.\bigskip

\textbf{F. Proof of Lemma 4}\medskip

Let us denote by $\nu^2_{k,n}(r)$ the power of the dominant path and $2\sigma^2_{k,n}(r)$ the power of the scattered paths of each link in LoS/NLoS at a ground distance $r$. The channel power gain of each link can be expressed in cartesian coordinates  as $|h_{k,n|r}|^2 = X^2_{k,n|r} + Y^2_{k,n|r}$, where $X_{k,n|r}$ and $Y_{k,n|r}$ are independent random variables which are normally distributed:
\begin{equation}
    X_{k,n|r}\sim \mathcal{N}\left(\nu_{k,n}(r) \cos \theta,\sigma^2_{k,n}(r)\right)\quad\mathrm{and}\quad Y_{k,n|r}\sim \mathcal{N}\left(\nu_{k,n}(r) \sin \theta,\sigma^2_{k,n}(r)\right), \quad \theta \in [0,2\pi[.
\end{equation}
By the properties of the normal distribution, it is known that $\sfrac{X^2_{k,n|r}}{\sigma^2_{k,n}(r)}$ and $\sfrac{Y^2_{k,n|r}}{\sigma^2_{k,n}(r)}$ follow a chi-square distribution with one degree of freedom, and therefore we have
\begin{alignat*}{3}
&\mathcal{L}_{X^2_{k,n|r}}(s) = \exp{-\frac{\nu^2_{k,n}(r)\cos^2\theta}{1+2\sigma^2_{k,n}(r)\: s}s}(1+2\sigma^2_{k,n}(r)\:s)^{-\frac{1}{2}} \\ 
&\mathcal{L}_{Y^2_{k,n|r}}(s) = \exp{-\frac{\nu^2_{k,n}(r)\sin^2\theta}{1+2\sigma^2_{k,n}(r)\: s}s}(1+2\sigma^2_{k,n}(r)\:s)^{-\frac{1}{2}}.
\end{alignat*}
Finally, Equation \eqref{eqn:LIT} is obtained knowing that $\mathcal{L}_{|h_{k,n|r}|^2}(s) = \mathcal{L}_{X^2_{k,n|r}}(s)\mathcal{L}_{Y^2_{k,n|r}}(s)$,
\begin{equation*}
    \nu^2_{k,n}(r) = \frac{K_{k,n}(r)}{K_{k,n}(r)+1}\Omega_{k,n}(r)\qquad\mathrm{and}\qquad 2\sigma^2_{k,n}(r) = \frac{1}{K_{k,n}(r)+1}\Omega_{k,n}(r),
\end{equation*}
where $\Omega_{k,n}(r) = 1$ with normalised power.
\bigskip

\textbf{F. Proof of Proposition 3}\medskip

The LT of $I_{k}(t)$ conditioned on $r_k(t)$ is given by
  \begin{align}
    \nonumber&\mathcal{L}_{I_{k}(t)|r_{k}(t)}(s) = \esp{-|r_{k}(t)}{\exp{-s\sum_{n=\{L,N\}}\sum_{\bm{\phi}_j \in \bm{\Phi}_{k,n,I}(t)} P_k G_{k,m} L_{k,n}^{-1}(\rVert \bm{\phi}_j \rVert) |h_{k,n|\rVert\bm{\phi}_j\rVert}|^2}}\\
    \nonumber&\overset{(a)}{=} \prod_{n=\{L,N\}}\esp{-|r_{k}(t)}{\prod_{\bm{\phi}_j \in \bm{\Phi}_{k,n,I}(t)}\exp{-s\: P_k G_{k,m} L_{k,n}^{-1}(\rVert \bm{\phi}_j \rVert)|h_{k,n|\rVert\bm{\phi}_j\rVert}|^2}}\\
    \nonumber&\overset{(b)}{=} \prod_{n=\{L,N\}}\esp{\bm{\Phi}_{k,I,n}(t)|r_{k}(t)}{\prod_{\bm{\phi}_j \in \bm{\Phi}_{k,n,I}(t)}\esp{|h_{k,n|\rVert\bm{\phi}_j\rVert}|^2}{\exp{-s\:P_k G_{k,m}L_{k,n}^{-1}(\rVert \bm{\phi}_j \rVert)|h_{k,n|\rVert\bm{\phi}_j\rVert}|^2}}}\\
     \nonumber&\overset{(c)}{=} \prod_{n=\{L,N\}} \exp{-2\pi \int_{0}^\infty \left(1 - \mathcal{L}_{|{h}_{k,n|r}|^2}\left( s \:  P_k G_{k,m} L_{k,n}^{-1}(r)\right) \right) p_{k,n}(r) \lambda_{k,I|r_k(t)}(r,t) \: r\mathrm{d}r}\\
     &= \exp{-2\pi\: \int_{0}^\infty \left(1 - \sum_{n=\{L,N\}} p_{k,n}(r) \mathcal{L}_{|{h}_{k,n|r}|^2}\left(s\: P_k G_{k,m} L_{k,n}^{-1}(r)\right) \right) \lambda_{k,I|r_k(t)}(r,t) r\mathrm{d}r}.
\end{align}
where $(a)$ is obtained since all the links are supposed to be independent. The PPs $\bm{\Phi}_{k,n,I}(t)$ are generated using an independent thinning of $\bm{\Phi}_{k,I}(t)$ with probability $p_{k,n}(r)$. The probability generating functional of a PPP in polar coordinates is used to obtain $(c)$.
\end{document}

%% file: Introduction.tex
The ease of deployment, the mobility, and the flying capability of Unmanned Aerial Vehicles (UAV) are strong arguments for their use as relays. They are not constrained by the ground topology, and can achieve high coverage since they communicate mostly via Line-of-Sight (LoS) links. For example, this is especially helpful in mmWaves scenarios (where short LoS links and large numbers of Relay Nodes (RN) are needed owing to the high path loss), or when the links with Terrestrial Base Stations (TBS) are obstructed. Additionally, their mobility enables to dynamically adjust their position to enhance the network performance.\smallskip

Relay networks have been extensively studied in the literature. As a non-exhaustive list, a mathematical framework for the analysis of two-hop relay-aided networks is provided in \cite{Lu2015}. A best biased average received power association is considered. In \cite{Belbase2018} and \cite{Biswas2016}, the performance of a two-hop link in a mmWave scenario is evaluated. The destination is at a fixed position, while there are either one unique or several sources. Additionally, two relay selection techniques are discussed in \cite{Biswas2016}. The developed framework is however only valid for terrestrial RNs, and can not be exploited for UAVs since the altitude and the mobility are not taken into account. 

Still in a mmWave scenario, UAV relay links are studied in the secrecy context of \cite{Ma2019}, in the presence of eavesdroppers. The performances of a cooperative UAV relay network are evaluated in \cite{Wang2019}, for a Non Orthogonal Multiple Access (NOMA) scenario involving two fixed users. The UAV relay number and locations are also fixed, but the interfering TBSs are modelled using a Homogeneous Poisson Point Process (HPPP). Also in a NOMA scenario, \cite{Li2020} develops a framework for satellite to ground User Equipment (UE) communications through UAV RNs thanks to the Decode-and-Forward (DF) protocol.  Most studies model the nodes with HPPPs, but \cite{Hayajneh2018,Ji2020} use a Poisson cluster process to generate fleets of UAVs and groups of UEs. This process enables to represent post-disaster scenarios and relay-assisted hotspots more realistically. Finally, the CP of a typical UE associated with UAV RNs or TBSs depending on the received average power is evaluated in \cite{newHybridAssoc}. It is computed in a finite circular area, using a binomial point process to model the static UAVs' positions. Additionally, LoS and Non Line-of-Sight (NLoS) propagation is considered for the backhaul link. \smallskip

The introduction of the mobility for UAV-based networks is evaluated in \cite{Mozaffari2016,Choi2020,Enayati2019,Sharma2019,Banagar2019}, with deterministic or stochastic trajectories, in two or three dimensions. In \cite{Mozaffari2016}, one UAV is flying at a fixed altitude around a circular area. In \cite{Choi2020}, a fleet of UAVs is periodically distributed on a line and flies in the orthogonal direction with a constant speed. \cite{Enayati2019} studies two families of stochastic trajectory processes providing an uniform coverage at each time step. In three dimensions, \cite{Sharma2019} considers a fixed number of UAVs uniformly deployed in a cylindrical area. The motion is characterised by a random waypoint model vertically, and a random walk horizontally. Finally, \cite{Banagar2019} considers UAVs moving in a random direction, or flying towards the users they serve. \smallskip

To our best knowledge, relay networks including both UAVs and mobility aspects have not been studied yet in a stochastic geometry framework with realistic propagation scenarios. 

\subsection*{Contributions}
The contributions of this paper are summarised here below:
\begin{enumerate}
    \item We consider a relay network composed of TBSs and UAV RNs in an infinite area, with positions modelled using two independent HPPPs. The UAV RNs are either hovering or are allowed to move following two mobility schemes presented in \cite{Banagar2019}. The association policy is the following: the typical UE is either served via a direct link (from a TBS), or via a relay link (TBS - UAV RN - UE). The selection of the link is performed using a nearest neighbour association rule. 
    \item We derive an approximate expression for the Coverage Probability (CP) of the typical UE, and the CP of each link taken independently. We resort to the Gil-Pelaez theorem \cite{GilPelaez} for an increased tractability.
    \item Different path loss and small-scale fading models are used depending on the considered link, in order to make the scenario more realistic. Every link is either in LoS or NLoS, based on a probability model developed for Air-to-Ground (A2G) and Ground-to-Ground (G2G) cases. The parameters of the fading and path loss of each link depend on its LoS/NLoS and A2G/G2G nature. Regarding the small-scale fading, Rician fading is considered, and the introduction of distance-dependent K factors is proposed.
\end{enumerate}
\smallskip

This paper is organised as follows: the system model is detailed in Section \ref{section:system_model}. Then, the coverage probability is derived in Section \ref{section:coverage_probability}. Finally, numerical results are presented in Section \ref{section:numerical_results}.

%% file: System_Model.tex
\begin{figure*}[b]
\rule{\linewidth}{0.5pt}
\begin{equation}
\lambda_{RD,I|r_{RD,0}}(r,t) = \left\{
\begin{alignedat}{10}
&\lambda_R \quad &&\mathrm{if}\quad && r_{RD,0} + vt \leq r\\
&\frac{\lambda_R}{\pi} \arccos\left(\frac{r_{RD,0}^2 - r^2 - v^2t^2}{2rvt}\right)\quad&& \mathrm{if}\quad && |r_{RD,0}-vt| \leq r < r_{RD,0} + vt\\
& \lambda_R \: \mathbbm{1} \left(r_{RD,0}< vt\right) \quad && \mathrm{if} \quad && 0 \leq r < |r_{RD,0}-vt|
\end{alignedat}
\right.
\label{eqn:densityMob} \tag{1}
\end{equation}
\end{figure*}

\label{section:system_model}
\subsection{Network topology}
We consider a typical UE located at $(0,0)$ in $\mathbb{R}^2$. The locations of the UAV RNs and TBSs are modelled as two independent HPPPs $\bm{\Phi}_R$ and $\bm{\Phi}_T$ of densities $\lambda_{R}$ and $\lambda_{T}$. The UAVs are all flying at a fixed altitude $H_R$, while the TBSs and the typical UE are at ground level. The transmit powers of the UAVs and TBSs are respectively given by $P_R$ and $P_T$. The network is assumed to be highly congested, such that all RNs and TBSs interfere with each other while transmitting.

\subsection{Mobility Schemes}
The UAVs are allowed to move following two mobility schemes in $\mathbb{R}^2$ presented in \cite{Banagar2019}, both at a fixed altitude.\smallskip

In the first mobility scheme, all the UAVs move independently at a constant speed $v$ in random directions in $\mathbb{R}^2$. It has been shown in \cite{Banagar2019} that the RNs' HPPP remains an HPPP with the same density at every time $t \geq 0$ with this scheme. Therefore, assuming an instantaneous handover between different RNs, the performance at the typical UE is not affected by the movements of the UAV RNs.\smallskip

In the second mobility scheme, all the UAVs move independently at a constant speed $v$ in random directions in $\mathbb{R}^2$, except the serving UAV, which moves towards the typical UE at position $(0,0)$ and then hovers above it. Knowing the initial distance between the selected RN and the typical UE, the density of the RNs interferers' Point Process (PP) is given by \eqref{eqn:densityMob} in that scenario \cite{Banagar2019}. With this mobility scheme, the distance between the typical UE and the selected RN at time $t\geq 0$ is given by 
\begin{equation}
    r_{RD}(t) = (r_{RD,0} - vt)\:u(r_{RD,0} - vt), \tag{2}
\end{equation}
with $r_{RD,0}=r_{RD}(0)$, and $u$ being the Heaviside step function.

\subsection{Association Rule}
For sake of mathematical tractability, the nearest neighbour criterion is used as association rule to select the serving nodes: 
\begin{equation}
    \tilde{\bm{\phi}} = \argmin_{\bm{\phi}_i \in \bm{\Phi}} \left\{\left\rVert \bm{\phi}_i - \bm{x} \right\rVert \right\}. \tag{3}
\end{equation}
This rule is also applied to make the selection between the best direct link (TBS to UE) and the best relay link (TBS to UE through a RN). Since the selected links can vary at different time instants with mobile UAVs, we assume instantaneous handovers between different UAVs or between direct and relay links.\smallskip

\begin{figure}[H]
    \centering
    \includegraphics[width=0.7\linewidth]{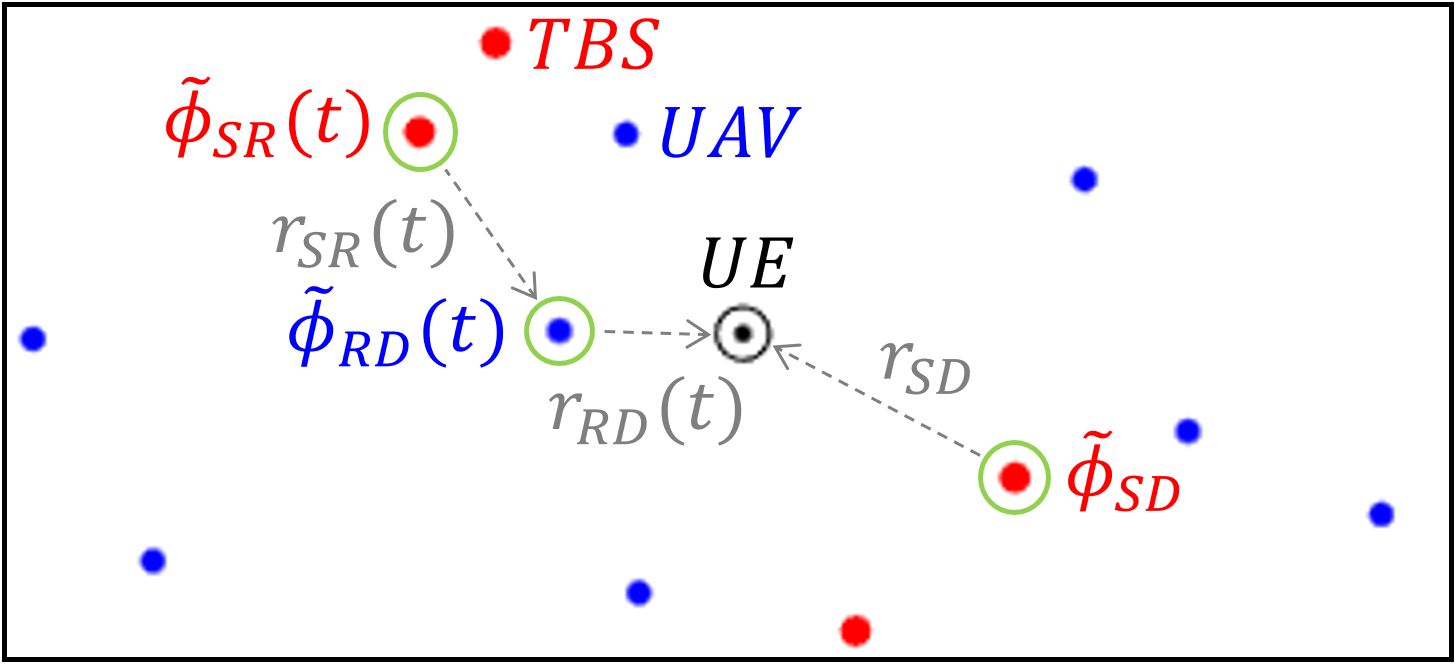}
    \caption{\small Illustration of the scenario: the typical UE is at $(0,0)$ and selects either the direct or the relay link depending on the distances between the nodes. The RNs and TBSs are respectively represented by blue and red dots. The selected nodes are circled in green.}
    \label{fig:scenario}
\end{figure}

In summary, the typical UE selects the direct link if the distance with its closest TBS $\tilde{\bm{\phi}}_{SD}$ is smaller than the distance with its closest RN $\tilde{\bm{\phi}}_{RD}(t)$. Otherwise the relay link is selected. In that case, the selected RN is served by its closest TBS $\tilde{\bm{\phi}}_{SR}(t)$. This scenario is illustrated in Figure \ref{fig:scenario}.

\vspace*{-0.3cm}

\subsection{Relay Transmission Protocol}
In this paper, the UAVs adopt the Decode-and-Forward (DF) protocol, executed over two time instants. At the first time instant, all the TBSs are transmitting to UEs or RNs. Thus the TBSs interfere with each other at that moment. At the second time instant, all the RNs re-transmit the signal to the UEs. Therefore, the RNs are interfering with each other at that instant. The Signal-to-Interference-plus-Noise-Ratio (SINR) of the relay link is given by \cite{Laneman2004}  
\begin{equation}
    \chi_{SRD}(t) = \min\left(\chi_{SR}(t),\chi_{RD}(t)\right),
    \label{eq:SINRDF} \tag{4}
\end{equation}
where $\chi_{SR}(t)$ is the SINR between the TBS and the RN, and $\chi_{RD}(t)$ is the SINR between the RN and the typical UE. 

\subsection{Beamforming model}
The TBSs and UAVs are equipped with antenna arrays to generate directional beams and perform transmit and receive beamforming. This leads to an SINR increase for the same transmitted power, enabling to compensate for a high path loss. As proposed in \cite{Belbase2018,Biswas2016,Ma2019}, a simple sectored antenna model is introduced:
\begin{equation}
    G(\psi,\theta) = \left\{
    \begin{matrix}
    G_{M} &\mathrm{if} |\psi| \leq \frac{\psi_{bw}}{2}\ \mathrm{and}\ |\theta| \leq \frac{\theta_{bw}}{2}\\
    G_{m} &\mathrm{otherwise}
    \end{matrix}
    \right., \tag{5}
\end{equation}
where $G_{M}$ is the maximum beamforming gain obtained when the elevation angle $\theta$ and the azimuth angle $\psi$ are respectively within the half elevation and azimuth power beamwidth $\theta_{bw}$ and $\psi_{bw}$, and $G_{m}$ is the minimum beamforming gain obtained in the other cases. In practice, it will be assumed that the beam of the target receiver is perfectly aligned with the beam of the transmitter and misaligned with the beams of the interferers, leading thereby to the array gains listed in Table \ref{tab:beamforming_gains}. The maximum and minimum beamforming gains of the TBSs and RNs are respectively denoted by $G_{TM}$, $G_{Tm}$, $G_{RM}$ and $G_{Rm}$.
\begin{table}[H]
\begin{center}
\begin{tabular}{|l||l|l|}
    \hline
         & \multicolumn{1}{c|}{Target} & \multicolumn{1}{c|}{Interference} \\\hline\hline
            {TBS-UE} & $G_{SD,M}=G_{TM}$ & $G_{SD,m}=G_{Tm}$ \\\hline
        {TBS-RN} & $G_{SR,M}=G_{TM}G_{RM}$ & $G_{SR,m}=G_{Tm}G_{Rm}$ \\\hline
        {RN-UE} & $G_{RD,M} = G_{RM}$ & $G_{RD,m} = G_{Rm}$ \\\hline
    \end{tabular}
    \caption{\small Beamforming gains for the different links.}
    \label{tab:beamforming_gains}
\end{center}
\end{table}
\vspace*{-0.6cm}
\subsection{Propagation model}
The channel model takes into account path loss and small-scale fading. The parameters of the different models are different for A2G and G2G links, in LoS or NLoS. We also introduce a model for the LoS probability.\smallskip

\paragraph{Path loss}
We consider two linear path loss models for A2G and G2G links \cite{Khawaja2019}: 
\begin{equation}
    L_{ln}(r)\ [dB] = \beta_{ln} + 10\:\alpha_{ln} \: \log_{10}\left(r\right), \tag{6}
\end{equation}
where $r$ is the distance between the two communicating nodes, $l = \{A,G\}$ for A2G or G2G links, and $n = \{L,N\}$ for links in LoS or NLoS. The parameters $\beta_{ln}$ are the floating intercepts while the parameters $\alpha_{ln}$ are the path loss exponents. In order to avoid convergence issues with the numerical integration of the expressions for G2G links, as in \cite{Khoshkholgh2019}, the distance $r$ is replaced by $1+r$ in the path loss model.\smallskip

\paragraph{Small-scale fading}
We assume that the propagation is affected by Rician fading. Knowing the ground distance $r$ (in two dimensions) between the two communicating nodes, it is modelled by a random variable $h_{ln|r}$, with normalised power and a distance-dependent K factor $K_{ln}(r)$, different for A2G/G2G links in LoS/NLoS. We suppose that the small-scale fading of different links are independent, and that the random variables at different time instants for the same link are independent.\smallskip

\paragraph{LoS probability (G2G links)} 
In this paper, the 3rd Generation Partnership Project (3GPP) $d_1/d_2$ model is selected, with a LoS probability given by \cite{Sun2015}
\begin{equation}
p_{GL}(r) = \min\left(\frac{d_1}{r},1\right) \left(1 - \exp{-\frac{r}{d_2}}\right) + \exp{-\frac{r}{d_2}}, \tag{7}
\end{equation}
where $r$ is the ground distance. The parameters $d_1$ and $d_2$ are respectively called the near-field and far-field critical distances.\smallskip

\paragraph{LoS probability (A2G links)} A commonly used LoS probability function for A2G links is given by \cite{Al-Hourani2014}
\begin{equation}
    p_{AL}(r) = \frac{1}{1 + a\  \exp{-b\left(\frac{180}{\pi}\arctan\left(\frac{H_R}{r}\right)-a\right)}},
    \label{eqn:LOSprobA2G} \tag{8}
\end{equation}
where $r$ is the ground distance, and $a$ and $b$ are two fitting parameters, directly linked to three other parameters with a physical meaning \cite{Al-Hourani2014}: the ratio of built-up land area to the total land area, the mean number of buildings per unit area, and a scale parameter that describes the buildings' heights distribution according to a Rayleigh Probability Density Function (PDF).\smallskip

\paragraph{Noise} We suppose that every link is affected by a white Gaussian noise of constant power $\sigma^2$.

%% file: Coverage_Probability.tex
In this section, an approximate mathematical expression for the CP of the typical UE is developed. For this purpose, three different links must be studied: the direct link from $\tilde{\bm{\phi}}_{SD}$ to the typical UE, the first hop of the relay link from $\tilde{\bm{\phi}}_{SR}(t)$ to $\tilde{\bm{\phi}}_{RD}(t)$, and the second hop of the relay link from $\tilde{\bm{\phi}}_{RD}(t)$ to the typical UE. The second mobility scheme is adopted. The results for the first mobility scheme or for hovering UAVs are deduced by setting $t=0$. Furthermore, the expressions obtained if the direct link is neglected (owing to a strong shadowing for example) can be easily deduced from this general framework.
\paragraph*{Notations} The symbol $\mathbb{P}_{X|Y}$ is the probability of $X$ conditioned on $Y$. The index $k = \{SD,SR,RD\}$ refers to the considered link, and the index $n = \{L,N\}$ refers to LoS/NLoS links. The direct link ($SD$) is a G2G link ($G$), while each hop of the relay link ($SR$ and $RD$) is an A2G link ($A$). Every time-varying quantity $Q(t)$ considered at time $t=0$ is denoted $Q_0$.

\subsection{SINR of each link}
Knowing the distance $r_k(t)$ between the two communicating nodes, the SINR of each link in LoS/NLoS can be defined as
\begin{equation}
    \chi_{k,n|r_{k}(t)}(t) = \frac{P_k G_{k,M} L^{-1}_{k,n}(r_k(t))|h_{k,n|r_k(t)}|^2}{I_{k}(t) + \sigma^2}, \tag{9}
\end{equation}
with $P_{SD} = P_{SR} = P_T$, $P_{RD} = P_R$, $h_{k,n|r_k(t)}$ being the small-scale fading of the considered link, and
\begin{align}
   & L_{SD,n}(r) = L_{Gn}(1+r), \nonumber\\
   & L_{SR,n}(r) = L_{RD,n}(r) = L_{An}\left(\sqrt{r^2+H_R^2}\right). \nonumber
\end{align}
The aggregate interferences of the three links are given by
\begin{equation}
I_k(t) = \sum_{\bm{\phi}_j \in \bm{\Phi}_{k,I}(t)} P_k G_{k,m} L_{k,n_j}^{-1}(\rVert \bm{\phi}_j \rVert)|h_{k,n_j|\rVert\bm{\phi}_j\rVert}|^2, \tag{10}
\end{equation}
with $n_j = \{L,N\}$ depending on whether the link between $\bm{\phi}_j$ and the node of interest is in LoS or in NLoS, and $h_{k,n_j,j|\rVert\bm{\phi}_j\rVert}$ being the small-scale fading of this link.\smallskip

The interferers' PPs are defined as 
\begin{equation}
\bm{\Phi}_{k,I}(t) = \bm{\Phi}_k \backslash \{\tilde{\bm{\phi}}_{k}(t)\},\tag{11}
\end{equation}
with $\bm{\Phi}_{SD} = \bm{\Phi}_{SR} = \bm{\Phi}_T$, and $\bm{\Phi}_{RD} = \bm{\Phi}_R$. Even if the selected TBS can vary at different time instants, with instantaneous handovers, \cite{Banagar2019} has shown that the distribution of $r_{SR}(t)$ is time-invariant.\smallskip

Knowing $r_{SD}$ and $r_{SR}(t)$, the densities of the interferers' PPs for the direct link and the first hop of the relay link are given by
\begin{alignat}{10}
&\lambda_{SD,I|r_{SD}}(r,t) &&=\lambda_T \: u(r-r_{SD}), \tag{12}\\
&\lambda_{SR,I|r_{SR}(t)}(r,t) &&= \lambda_T \: u(r-r_{SR}(t)). \tag{13}
\end{alignat}

\begin{figure*}[b]
\rule{\linewidth}{0.5pt}
\begin{multline}
  \mathcal{A}_{SRD} = \exp{-\pi\lambda_T H_R^2}\left(1-\exp{-\pi\lambda_R\: v^2t^2}\right) \\+ 2\pi\lambda_R\:\exp{-\pi\lambda_T(H_R^2+v^2t^2)}\int_{vt}^\infty r\:\exp{-\pi(\lambda_R+\lambda_T)r^2 + 2\pi\lambda_T vt\:r}\mathrm{d}r\label{eqn:probsel} \tag{18}
\end{multline}
\end{figure*}
\vspace*{-0.0cm}

\subsection{PDF of the initial distances between the selected nodes}
Since the TBSs' and UAVs' HPPPs are independent, $r_{SD}$ and $r_{RD}(t)$ are independent. It is also the case for $r_{SR}(t)$ and $r_{RD}(t)$. However, $r_{SR}(t)$ is dependent of $r_{SD}$ since there is a non zero probability for the same TBS to be selected by the typical UE and the selected RN, as illustrated by Lemma 1.
\medskip

\textit{\textbf{Lemma 1:} The probability for the same TBS to be selected by the typical UE and the selected RN, conditioned on $r_{SD}$ and $r_{RD,0}$ is given by
\begin{align}
    &\Pr{\bm{\tilde{\phi}}_{SD},\bm{\tilde{\phi}}_{SR}(t)|r_{SD},r_{RD,0}}{\bm{\tilde{\phi}}_{SD} = \bm{\tilde{\phi}}_{SR}(t)}\label{eqn:probOfdependency} \tag{14} \\ 
    &= \exp{-\pi\lambda_T\left(r_{SD}^2+r_{RD}^2(t)\right)}I_0\left(2\pi\lambda_T\:r_{SD}r_{RD}(t)\right). \nonumber
\end{align}}
\textit{Proof:} see Appendix A. \hfill{\small$\blacksquare$}\medskip

Therefore, in order to simplify the mathematical expressions, Approximation 1 is introduced.\medskip

\textit{\textbf{Approximation 1:} We assume that $r_{SR}(t)$ is independent with $r_{SD}$.}
\medskip

Since \eqref{eqn:probOfdependency} decreases with $\lambda_T$, Approximation 1 becomes more accurate when the TBS' density increases because the dependence between $\tilde{\bm{\phi}}_{SD}$ and $\tilde{\bm{\phi}}_{SR}(t)$ decreases.\medskip

\textit{\textbf{Lemma 2:} With the nearest neighbour association rule, the PDFs of the initial distances between the couple of nodes of the first hop, second hop and direct link are given by
\begin{equation}
   f_{r_{k,0}}(r_{k,0}) = 2\pi\lambda_k\:r_{k,0}\: \exp{-\pi\lambda_k r_{k,0}^2},\label{eqn:PDFrSD} \tag{15}
\end{equation}
with $\lambda_{SD} = \lambda_{SR} = \lambda_T,\ \lambda_{RD} = \lambda_R$.}\smallskip

\textit{Proof:} see Appendix B. \hfill{\small$\blacksquare$}\medskip

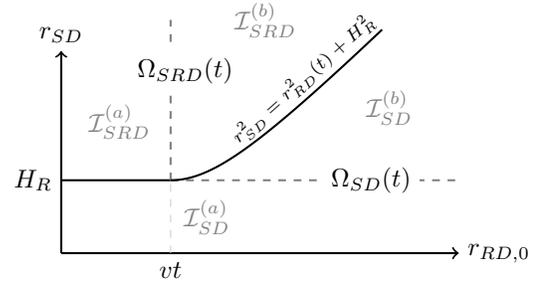
\begin{figure}[H]
\begin{center}
\begin{adjustbox}{width = 0.8\linewidth}
\begin{tikzpicture}
    \node (RSD) at (1,4) {$r_{SD}$};
    \node (RRD) at (7,1) {$r_{RD,0}$};
    \node[anchor=north] (vt) at (2.5,1) {$vt$};
    \node[anchor=east] (HR)  at (1,2) {$H_R$};
    \draw[->,thick] (1,1) -- (RSD);
    \draw[->,thick] (1,1) -- (RRD);	
    \draw[thick,-] (HR) -- (2.5,2);
    \draw[-,black!20!white,dashed] (vt) -- (2.5,2);
    \draw[thick,-,dashed,black!50!white] (2.5,2) -- (2.5,4.2);
    \draw[thick,-,dashed,black!50!white] (2.5,2) -- (6.5,2);
    \node[fill=white] (SD)  at (5.25,2) {$\Omega_{SD}(t)$};
    \node[black!50!white] (ISDA) at (3,1.5) {$\mathcal{I}_{SD}^{(a)}$};
    \node[black!50!white] (ISDB) at (5.5,3) {$\mathcal{I}_{SD}^{(b)}$};
    \node[fill=white] (SRD) at (2.7,3.5) {$\Omega_{SRD}(t)$};
    \node[black!50!white] (ISRDA) at (1.8,2.8) {$\mathcal{I}_{SRD}^{(a)}$};
    \node[black!50!white] (ISRDB) at (3.8,4.2) {$\mathcal{I}_{SRD}^{(b)}$};
    \draw[thick,samples=100,domain=0:2.9,shift={(2.5,2)},
    postaction={decorate,decoration={text along path,text align={right},raise=1ex,text={{\scriptsize $r_{SD}^2=r_{RD}^2(t)+H_R^2$}{}}}}] plot ({\x}, {sqrt(\x^2+1)-1});
\end{tikzpicture}
\end{adjustbox}
\end{center}
\caption{\small Regions of association with the nearest neighbour association rule for the second mobility scheme at time $t\geq 0$. The typical UE selects the direct or the relay link if the couple $(r_{SD},r_{RD,0})$ is respectively in $\Omega_{SD}(t)$ or $\Omega_{SRD}(t)$.}
\label{fig:associationRegionsNNAMob}
\end{figure}

With the nearest neighbour association rule, we define the regions of association $\Omega_{SD}(t)$ and $\Omega_{SRD}(t)$ as
\begin{alignat}{10}
    &\Omega_{SD}(t) &&:= \left\{(r_{SD},r_{RD,0})\:|\: r_{SD}^2 \leq r_{RD}^2(t) + H_R^2\right\}, \tag{16}\\
    &\Omega_{SRD}(t) &&:= \mathbb{R}_+^2\:\backslash\:\Omega_{SD}(t). \tag{17}
\end{alignat}
Therefore, at time $t\geq 0$, for a given couple of initial distances $\bm{x}=(r_{SD},r_{RD,0})$, the typical UE selects the direct or relay link respectively if $\bm{x}\in\Omega_{SD}(t)$ or $\bm{x}\in\Omega_{SRD}(t)$. This is illustrated in Figure \ref{fig:associationRegionsNNAMob}. The integration of the PDFs on these regions gives the probability of association with the direct and relay link. The result is presented in Lemma 3.\medskip

\begin{figure*}[b]
\rule{\linewidth}{0.5pt}
\begin{alignat}{2}
&\mathcal{I}_{SD}^{(a)}(\beta) &&= \int_0^{H_R}\mathcal{P}_{SD|r_{SD}=r}(\beta)f_{r_{SD}}(r) \:\mathrm{d}r\label{eqn:I1}\tag{20}\\
&\mathcal{I}_{SD}^{(b)}(\beta,t) &&= \exp{-\pi\lambda_R \left(v^2t^2-H_R^2\right)} \int_{H_R}^\infty \mathcal{P}_{SD|r_{SD}=r}(\beta) f_{r_{SD}}(r)\: \exp{-\pi\lambda_R \left(r^2 + 2vt\sqrt{r^2-H_R^2}\right)}\mathrm{d}r\label{eqn:I2}\tag{21}\\
&\mathcal{I}_{SRD}^{(a)}(\beta,t) &&= \exp{-\pi\lambda_T H_R^2}\int_0^{\infty} \mathcal{P}_{SR|r_{SR}(t)=r}(\beta) f_{r_{SR}(t)}(r) \: \mathrm{d}r\int_{0}^{vt} \mathcal{P}_{RD|r_{RD,0}=r}(\beta,t) f_{r_{RD,0}}(r)\:\mathrm{d}r\label{eqn:I3}\tag{22}\\
\nonumber&\mathcal{I}_{SRD}^{(b)}(\beta,t) &&= \exp{-\pi\lambda_T \left(H_R^2+v^2t^2\right)}\int_0^{\infty} \mathcal{P}_{SR|r_{SR}(t)=r}(\beta) f_{r_{SR}(t)}(r) \: \mathrm{d}r\\
&&&\qquad\qquad\qquad\qquad\qquad\qquad\qquad\cdot\int_{vt}^\infty \mathcal{P}_{RD|r_{RD,0}=r}(\beta,t)f_{r_{RD,0}}(r)\:\exp{-\pi\lambda_T (r^2-2vt\:r)}\mathrm{d}r\label{eqn:I4}\tag{23}
\end{alignat}
\rule{\linewidth}{0.5pt}
\begin{equation}
    \mathcal{P}_{k|n,r_{k}(t)}(\beta,t) = \frac{1}{2} + 
    \frac{1}{\pi}\int_0^\infty\frac{1}{\tau} 
    \mathrm{Im}\left[\mathcal{L}_{|h_{k,n|r_k(t)}|^2}\left(-j\tau P_kG_{k,M}L_{k,n}^{-1}(r_k(t))\right)
    \mathcal{L}_{I_{k}(t)|r_{k}(t)}(j\tau\beta) \: \exp{-j\tau\beta \sigma^2}\right]\mathrm{d}\tau\label{eqn:PSDRSD} \tag{25}
\end{equation}
\rule{\linewidth}{0.5pt}
\begin{equation}
    \mathcal{L}_{I_k(t)|r_k(t)}(s) = \exp{-2\pi \int_{0}^\infty \left(1 - \sum_{n=\{L,N\}} p_{k,n}(r) \mathcal{L}_{|{h}_{k,n|r}|^2}\left(s\:P_k G_{k,m}L_{k,n}^{-1}(r) \right)\right) \lambda_{k,I|r_k(t)}(r,t)\: r\mathrm{d}r} \label{eqn:LIT} \tag{27}
\end{equation}
\end{figure*}

\textit{\textbf{Lemma 3:} With the nearest neighbour association rule, the probability $\mathcal{A}_{SRD}$ that the typical UE selects the relay link at time $t\geq 0$ is given by \eqref{eqn:probsel}.}
\vspace*{0.1cm}

\textit{Proof:} see Appendix C. \hfill{\small$\blacksquare$}\medskip

\subsection{Coverage Probability}
Using the definition of the SINRs of each link, Lemmas 2 and Approximation 1, Proposition 1 gives the mathematical expression of the CP of the typical UE.
\medskip 

\textit{\textbf{Proposition 1:} With the second mobility scheme, considering the nearest neighbour association rule and selection between direct and relay link, the approximate CP of the typical UE at time $t\geq 0$ for a given SINR threshold $\beta$ is given by 
\begin{equation}
\mathcal{P}(\beta,t) = \mathcal{I}_{SD}^{(a)}(\beta) + \mathcal{I}_{SD}^{(b)}(\beta,t) + \mathcal{I}_{SRD}^{(a)}(\beta,t) + \mathcal{I}_{SRD}^{(b)}(\beta,t), \tag{19}
\end{equation}
where $\mathcal{I}_{SD}^{(a)}$, $\mathcal{I}_{SD}^{(b)}$, $\mathcal{I}_{SRD}^{(a)}$ and $\mathcal{I}_{SRD}^{(b)}$ are respectively given by \eqref{eqn:I1}-\eqref{eqn:I4}. The subscripts $SD$ and $SRD$ correspond to whether the considered couples $(r_{SD},r_{RD,0})$ are included in $\Omega_{SD}(t)$ or $\Omega_{SRD}(t)$, as shown in Figure \ref{fig:associationRegionsNNAMob}. In these expressions, $\mathcal{P}_{k|r_{k}(t)}$ are respectively the CP of each link conditioned on the distance between the two nodes of interest at time $t$.}\smallskip

\textit{Proof:} see Appendix D. \hfill{\small$\blacksquare$}\medskip

Moreover, the CPs of every link taken independently are obtained by removing the conditioning of $\mathcal{P}_{k|r_{k}(t)}$. Using the law of total probability, we have
\begin{equation}
    \mathcal{P}_{k|r_{k}(t)}(\beta,t) = \sum_{n=\{L,N\}}p_{k,n}(r_k(t))\:\mathcal{P}_{k|n,r_k(t)}(\beta,t). \tag{24}
\end{equation}
These expressions are functions of the CPs of each hop if these links are in LoS or NLoS, and respectively conditioned on the distances between the nodes of interest. The Gil-Pelaez theorem enables to compute these quantities based on their characteristic functions \cite{GilPelaez}. It enables to reduce the mathematical complexity of the framework (no multiple numerical derivations of the Laplace Transforms (LT) of the aggregate interferences for each link) using numerical integrations. Based on this theorem, Proposition 2 gives the expressions of $\mathcal{P}_{k|n,r_{k}(t)}$.
\medskip

\textit{\textbf{Proposition 2:} The CPs of each link in LoS or NLoS, conditioned on the distance between the two communicating nodes are given by \eqref{eqn:PSDRSD}. In these expressions, $\mathcal{L}_{|h_{k,n|r_k(t)}|^2}$ are the LTs of the small-scale fading channel power gain of each link, while $\mathcal{L}_{I_{k}(t)|r_{k}(t)}$ are the LTs of the aggregate interferences respectively conditioned on $r_{SD}$, $r_{SR}(t)$ and $r_{RD}(t)$.}\smallskip

\textit{Proof:} see Appendix E. \hfill{\small$\blacksquare$}\medskip

\subsection{LTs of the small-scale fading channel power gains}
\textit{\textbf{Lemma 4:} Under Rician fading, the LTs of the small-scale fading channel power gains of each link are given by
\begin{equation}
\mathcal{L}_{|h_{k,n|r}|^2}(s) = \frac{K_{k,n}(r)+1}{K_{k,n}(r)+1+s}\:\exp{-\frac{K_{k,n}(r)\:s}{K_{k,n}(r)+1+s}}.
\label{eqn:LTRicianFading} \tag{26}
\end{equation} 
} \smallskip
\textit{Proof:} see Appendix F. \hfill{\small$\blacksquare$}\medskip

\subsection{LTs of the aggregate interferences}
\textit{\textbf{Proposition 3:} The LTs of the aggregate interferences of each link at time $t \geq 0$, respectively conditioned on $r_{SD}$, $r_{SR}(t)$ and $r_{RD}(t)$ are given by \eqref{eqn:LIT}.}\smallskip

\textit{Proof:} see Appendix G. \hfill{\small$\blacksquare$}\medskip

%% file: Numerical_Results.tex
\label{section:numerical_results}

In this section, the CP of a typical UE for the parameters given in Table \ref{tab:sim_parameters} is evaluated using the developed framework. The impact of the introduction of mobile UAV RNs is studied in particular. We used constant values for the Rician fading K factors. The extension of the numerical results with realistic distance-dependent K factors introduced in the proposed framework is left for future studies.\smallskip

\begin{table}[t]
\hspace*{-0.3cm}
\begin{tabular}{c@{\hskip 0.05cm}c}
\begin{tabular}{|c|c|}
    \hline
    \multicolumn{2}{|c|}{G2G Links} \\\hline\hline
    $A_{GL}$ & 0.01 \\
    $A_{GN}$ & 0.01 \\
    $\alpha_{GL}$ & 3  \\
    $\alpha_{GN}$ & 4  \\
    $K_{GL}$ & 10 \\
    $K_{GN}$ & 0 \\
    $d_1$ & 18 $m$ (from \cite{Sun2015})\\
    $d_2$ & 63 $m$ (from \cite{Sun2015})\\\hline\hline
    \multicolumn{2}{|c|}{TBSs} \\\hline\hline
         $P_T$ & 1 $W$  \\
     $\lambda_T$ & $5 \cdot 10^{-8}$ $m^{-2}$ \\
     $G_{TM}$ & 2 \\
     $G_{Tm}$ & 0.5\\\hline\hline
     \multicolumn{2}{|c|}{MC simulations}\\\hline\hline
    Iterations & $50000$ \\
    Circular area radius & 100 $km$ \\\hline
\end{tabular}
&
\begin{tabular}{|c|c|}
    \hline
    \multicolumn{2}{|c|}{A2G Links} \\\hline\hline
    $A_{AL}$ & 0.01 \\
    $A_{AN}$ & 0.01 \\
    $\alpha_{AL}$ & 3  \\
    $\alpha_{AN}$ & 4  \\
    $K_{AL}$ & 10 \\
    $K_{AN}$ & 0 \\
    $a$ & 9.612 (from \cite{Al-Hourani2014})\\
    $b$ & 0.158 (from \cite{Al-Hourani2014})\\\hline\hline
    \multicolumn{2}{|c|}{UAV RNs}\\\hline\hline
     $P_R$ & 1 $W$  \\
     $\lambda_R$ & $10^{-7}$ $m^{-2}$\\
     $H_R$ & $100-2000$ $m$\\
     $G_{RM}$ & 1 \\
     $G_{Rm}$ & 1\\\hline\hline
     \multicolumn{2}{|c|}{Noise Variance} \\\hline\hline
    $\sigma^2$ & $10^{-10}$ W \\\hline
\end{tabular}
\\
\end{tabular}
    \caption{\small Network parameters.}
    \label{tab:sim_parameters}
\end{table}

\begin{figure}[t]
\vspace*{-0.3cm}
\begin{tikzpicture}
\node (fig1) at (0,0) {\includegraphics[width=\linewidth]{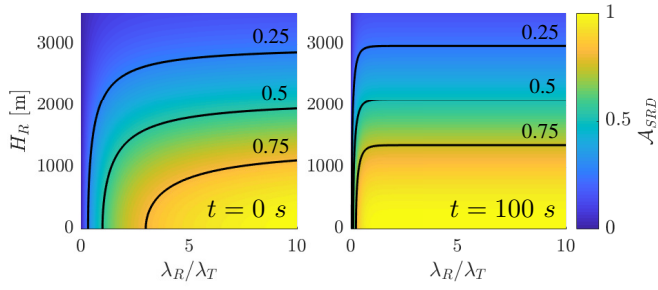}};
\node (251) at (-0.78,1.40) {{\footnotesize 0.25}};
\node (501) at (-0.73,0.64) {{\footnotesize 0.5}};
\node (751) at (-0.78,-0.06) {{\footnotesize 0.75}};
\node (252) at (2.8,1.45) {{\footnotesize 0.25}};
\node (502) at (2.83,0.73) {{\footnotesize 0.5}};
\node (752) at (2.8,0.14) {{\footnotesize 0.75}};
\node (t0)  at (-1.1,-0.9) {{$t=0\ s$}};
\node (t100)  at (2.3,-0.9) {{$t=100\ s$}};
\end{tikzpicture}
\caption{\small Probability of association with the relay link, at time \\$t=0$ $s$ (left) and $t =100$ $s$ (right), with $v=40$ $\sfrac{m}{s}$.}
\label{fig:probassoc}
\end{figure}

First, Figure \ref{fig:probassoc} shows the probability of association with a UAV for different RNs' altitudes and densities with the nearest neighbour association rule. The relay is often selected when the RNs’ density increases, but when this density is large compared to the TBSs’ density, the RNs’ altitude has the largest impact. Additionally, with the second mobility scheme, the impact of the RNs' density decreases when stationarity is reached. Indeed, since the selected RN moves toward the typical UE, the probability for the typical UE to select the relay link increases until the selected RN hovers above the typical UE. At that moment, the RNs' altitude is the only parameter to influence the probability of association. 

\begin{figure}[b]
\includegraphics[width=\linewidth]{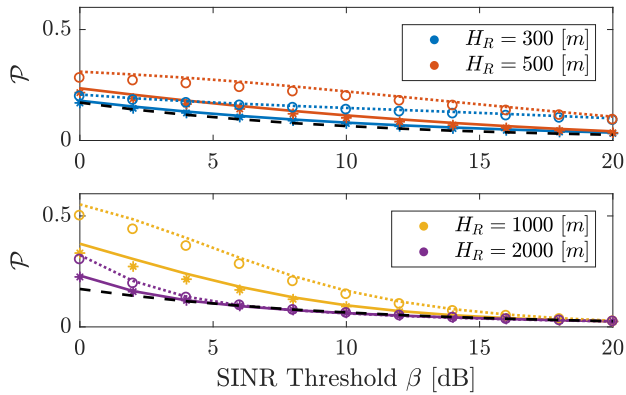}
\caption{\small CP of the typical UE for different RNs' altitudes at $t = 0\ s$ (solid lines) and $t = 2.5\bar{T}\ s$ (dotted lines) compared to the CP of the direct link (dashed lines). The expected travel time $\bar{T}$ is defined as $\bar{T} = \sfrac{\esp{}{r_{RD,0}}}{v}$.}
\label{fig:CP}
\end{figure}

Next, Figure \ref{fig:CP} illustrates the CP of the typical UE for different SINR thresholds, RNs' altitude and time. The LoS probability of the A2G links increases with the RNs' density and altitude following \eqref{eqn:LOSprobA2G}. This enables to reach higher power levels for the useful links and improve the SINR of the relay link. However, the power levels from the interferers of the first and second hop also increase. Additionally, when the RNs’ altitude increases, the distances between the typical UE and the RNs which are close tend all to be equal to the RNs’ altitude, degrading the SINR of the first and second hop. Therefore, in order to achieve the best relay link's CP at a given SINR threshold, the RNs' altitude and density must be jointly optimised, taking into account the association probability with the direct link when they are not neglected. For this network configuration, with a RNs' density of $10^{-7}$ $m^{-2}$, the relay link helps to improve the CP of the typical UE for RNs' altitudes higher than 300 meters. In that particular case, a low RNs' density limits the interference from the interfering RNs, while a high altitude enables to increase the LoS probability of both relay link's hops. For example, the best CP is achieved at RNs' altitudes close to 1000 meters for a SINR threshold of 0 dB with this network configuration.

\begin{figure}[H]
\includegraphics[width=\linewidth]{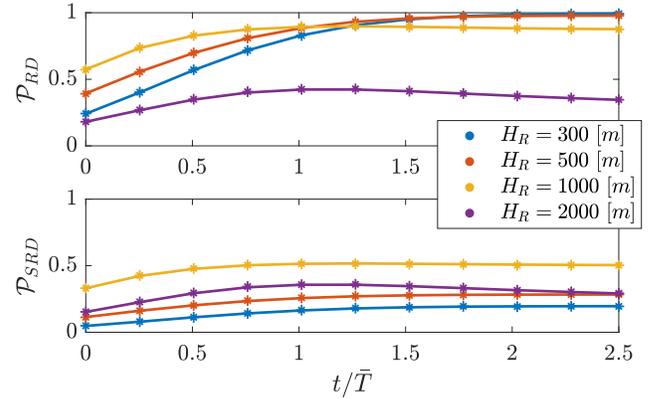}
\caption{\small Time evolution of the CP of the second hop and of the two hops at different RNs' altitudes for a SINR threshold $\beta = 0$ dB. The expected travel time $\bar{T}$ is defined as $\bar{T} = \sfrac{\esp{}{r_{RD,0}}}{v}$.}
\label{fig:CP_time}
\end{figure}

Finally, Figure \ref{fig:CP_time} shows the evolution of the CP of the second hop and the two hops with time for a fixed SINR threshold. The introduction of the second mobility scheme enables to further improve the CP of the typical UE. Since the selected RN is moving towards the typical UE, the LoS probability of the link increases to one. This leads to a large increase for the CP of the second hop. However, the CP of the first hop is time-invariant, therefore a smaller improvement is visible for the total CP of the relay link. At low RNs' altitudes, the CP of the first hop is a bottleneck which limits considerably the achievable coverage. Owing to the lower TBSs' density, the useful link of the first hop is mostly in NLoS. Contrariwise, at high RNs' altitudes, the interfering power is increased owing to the higher probability for the A2G links to be in LoS. In that case, the second hop is limiting the coverage of the relay link. Moreover, the best CP is not achieved at stationarity, but it is reached at a time close to the expected travel time defined as $\bar{T} = \sfrac{\esp{}{r_{RD,0}}}{v}$. This is caused by the increase of the interferers' density close to the typical UE for $t > \bar{T}$ , as shown in \cite{Banagar2019}. This is especially visible for high RNs' altitudes since the interfering power is higher. For large RNs' densities, the CP' improvement is still visible, but less significant since the number of interferers grows proportionally.

%% file: Conclusion.tex
In this paper, a framework integrating UAV as relays in communication networks has been developed using stochastic geometry. UAVs either hover at a fixed position, or are allowed to fly according to two mobility schemes. It has been shown for the considered network configuration that a joint optimisation of the RNs' density and altitude enables to enhance the CP of the typical UE thanks to relay links. Additionally, relay links can provide a good alternative to direct links in case of severe obstruction. The second mobility scheme enables to further improve the overall performance, thanks to a higher probability for the second hop's link to be in LoS, without affecting the performance of the first hop.\smallskip 

The current system model could be improved according to the following directions: first, the nearest neighbour criterion could be replaced by the maximum average power criterion. Indeed, the latter does not take the nature of the link into account (G2G/A2G, LoS/NLoS). In addition, more realistic distance-dependent K factor functions could be introduced based on measurement campaigns or COST models. Moreover, a tractable approach to include shadowing must be investigated. For example, the $\kappa - \mu$ and $\eta - \mu$ channel models presented in \cite{kappamuetamu} are other alternatives for the channel modelling. 